\documentclass[aps,prb,groupedaddress]{revtex4}
\usepackage{dcolumn}
\usepackage{bm}
\usepackage{amsfonts}
\usepackage{latexsym}
\usepackage{graphicx}

\newcommand{\be}{\begin{equation}}
\newcommand{\ee}{\end{equation}}
\newcommand{\bea}{\begin{eqnarray}}
\newcommand{\eea}{\end{eqnarray}}

\newcommand{\figwidth}{0.4\columnwidth}
\newcommand{\figwidths}{0.425\columnwidth}

\begin{document}


\title{Zero Temperature Series Expansions for the Kondo Lattice Model}


\author{Weihong Zheng}
\email[]{w.zheng@unsw.edu.au}
\homepage[]{http://www.phys.unsw.edu.au/~zwh}
\affiliation{School of Physics,
The University of New South Wales,
Sydney, NSW 2052, Australia.}


\author{J. Oitmaa}
\email[]{j.oitmaa@unsw.edu.au}
\affiliation{School of Physics,
The University of New South Wales,
Sydney, NSW 2052, Australia.}

%

\date{\today}

\begin{abstract}
We present new results for the Kondo lattice model of strongly correlated
electrons, in 1-, 2-, and 3-dimensions, obtained from high-order
linked-cluster series expansions. Results are given for varies
ground state properties at half-filling, and for spin and charge
excitations. The existence and nature of the predicted quantum phase
transition are explored.
\end{abstract}


\maketitle

\section{\label{sec:intro}INTRODUCTION}

The Kondo lattice model, described by the usual Hamiltonian
\be
H = -t \sum_{\langle ij\rangle \sigma } ( c_{i\sigma}^{\dag} c_{j\sigma} + h.c.) + J \sum_i {\bf S}_i \cdot {\bf s}_i
\ee
represents a band of conduction electrons, interacting via a spin-exchange term with
a set of immobile $s=\frac{1}{2}$ spins ${\bf S}_i$ (f electrons).

The model has been extensively studied in connection with a class of materials known as
``Kondo insulators"\cite{aep92} ($J>0$), and in connection with the manganites ($J<0$)\cite{dag98}.
Despite the apparent simplicity of the model, in which neither the conduction
electrons nor localized spins interact directly among themselves, the spin exchange leads to a
strongly-correlated many-body system. No exact results are known for either
ground state or thermodynamic properties for general $J/t$, in any dimension.

The model incorporates two competing physical processes. 
In the strong-coupling (large $\vert J\vert$)  
limit, the conduction electrons are ``frozen out" via the formation of local singlets
($J>0$) or triplets ($J<0$). In either case there will be a gap to spin excitations and spin
correlations will be short ranged. On the other hand, at weak coupling, the
conduction electrons can induce the usual RKKY interaction between localized spins, giving
rise to possible magnetically ordered phases with no spin gap and long-range correlations.
In one dimension there will be a smooth crossover from large $\vert J\vert$ to small
$\vert J\vert $ behaviour, but in higher dimension a quantum phase transition is expected.

A great deal of work has been carried out on the one dimensional model,
using a variety of analytic and numerical methods, and we refer the reader
to a recent review\cite{tsu97}. In higher dimension there have been
mean-field approaches\cite{lac79,zha00,jur01}, quantum Monte Carlo
calculations\cite{ass99}, and a series expansion study\cite{shi95}.
These studies, which are all for the half-filled case, conclude
that a quantum phase transition, at which the spin gap vanishes
continuously, occurs at $(J/t)_c \simeq 1.45\pm 0.05$ in the
2D $J>0$ case, while Refs. \onlinecite{jur01,shi95} give
$(J/t)_c \simeq 1.833, 2.0$ respectively for the 3D $J>0$ case.
There have not been, to our knowledge, any similar studies for the
case of ferromagnetic coupling.

Our aim in this paper is to study the Kondo lattice model in
1, 2 and 3-dimensions via series expansion methods. We have considerably
extended the calculations of Ref. \onlinecite{shi95}, by obtaining
longer series, by using also expansions about the Ising limit, and by studying 
also the energies of elementary excitations.

Linked-cluster series expansions have been used successfully for many years
to study strongly interacting lattice models. A recent review\cite{gelfand4}
describes the basic approach and some of the results which have been obtained.
The method is applicable in any dimension, is particularly suited to
locating critical points and is free from finite size corrections or
minus sign problems which hamper other numerical approaches.
On the other hand good convergence may be limited to particular
regions of the phase diagram.

The Hamiltonian is written in the generic form 
$H=H_0+\lambda V$
where $H_0$ has a simple known ground state. The remaining term(s)
in $H$ are treated perturbatively, to high order. In this way
the ground state energy, correlations, susceptibilities, etc,
are expressed as power series in $\lambda$. These are then analysed
by standard methods\cite{gut}. An extension of the basic linked-cluster
method\cite{gel96,gelfand4} allows the computation of the full
dispersion relation for elementary excitations, which can yield
energy gaps.

\begin{figure}[b]
  { \centering
    \includegraphics[width=\figwidth]{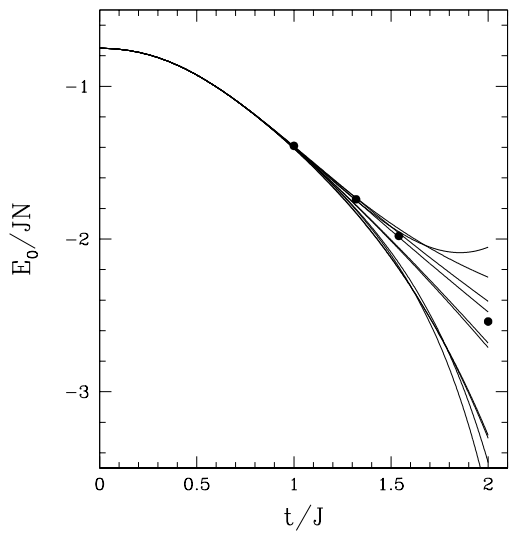}
    \caption{The ground state energy for the 1D Kondo lattice model. The solid 
    points are the DMRG results\cite{yu93}.
    \label{fig_1d_e0}} 
  }
\end{figure} 

For the present model the simplest choice is to take 
$H_0 = J \sum_i {\bf S}_i \cdot {\bf s}_i$, a sum of single-site
exchange terms. The unperturbed ground state is then a simple
product state of dimer states. This is the approached used in 
Ref. \onlinecite{shi95}, and also our first method here.
We refer to these as ``dimer expansions". An alternative is to write
the exchange term as
\be
J \sum_i [  S^z_i  s^z_i + \frac{1}{2} (S_i^+ s_i^- + S_i^- s_i^+ ) ]
\ee
and to take only the first term as $H_0$, In such an ``Ising expansion"
both the spin-fluctuation and hopping terms are
treated perturbatively.
To remove the degeneracy in $H_0$,
we  add following two terms for conduction spins into $H_0$
\be
 J' \sum_{\langle ij\rangle} ( s_i^z s_j^z + 1/4 ) + h \sum_i [ (-1)^i s^z_i + 1/2 ]
\ee
and subtract them from the perturbation term, so the overall Hamiltonian is
\begin{widetext}
\bea
H &=& H_0 + \lambda V  \\
H_0 &=&   J \sum_i ( S_i^z s_i^z  )
     + J' \sum_{\langle ij\rangle } ( s_i^z s_j^z + 1/4 )
     + h \sum_i [ (-1)^i s^z_i + 1/2 ]  \\
 V  &=&  J \sum_i ( S_i^x s_i^x + S_i^y s_i^y )
    - J' \sum_{\langle ij\rangle} ( s_i^z s_j^z +1/4 )
   - h/2 \sum_i [(-1)^i s^z_i + 1/2]
     - t \sum_{\langle ij\rangle\sigma} ( c^+_{i \sigma} c_{j \sigma} + h.c. )
\eea
\end{widetext}

Series in power of $\lambda$ are computed for given values of $J,t,J'$ and $h$,
and extrapolated to $\lambda=1$ where the original Hamiltonian is recovered.
Such expansions are appropriate for magnetically ordered phases, although they can
also yield accurate results in other cases.

In the remainder of this paper we will present and discuss our results for
the 1D case (Section II), for the 2D square lattice  and
for the 3D simple cubic lattice (Section III). An overall summary is given at the
end.

\section{\label{sec:1d}The 1D Kondo Lattice Model}
\begin{figure*}[tb]
  { \centering
    \includegraphics[width=\figwidths]{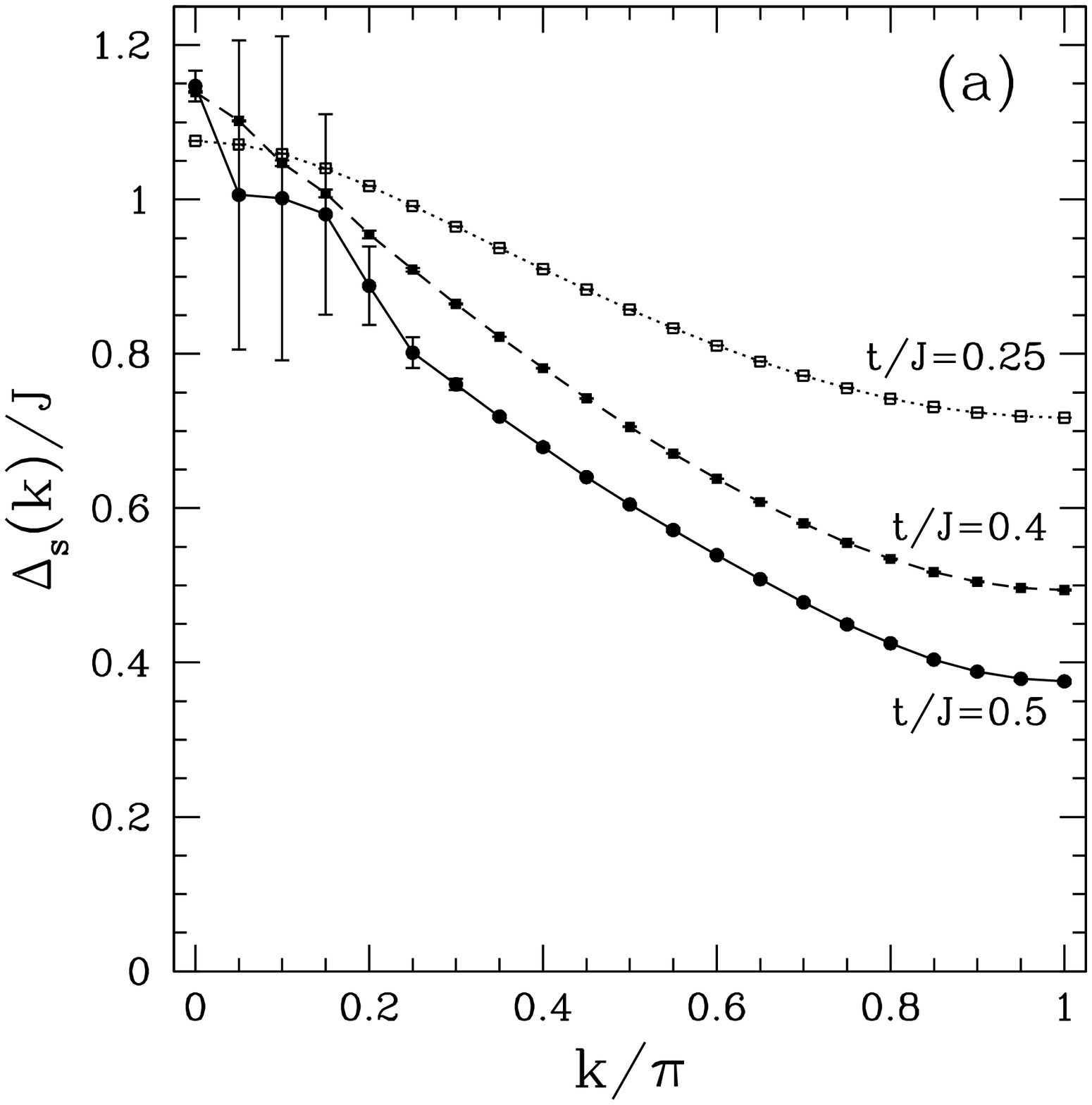}\hspace{1cm}
    \includegraphics[width=\figwidths]{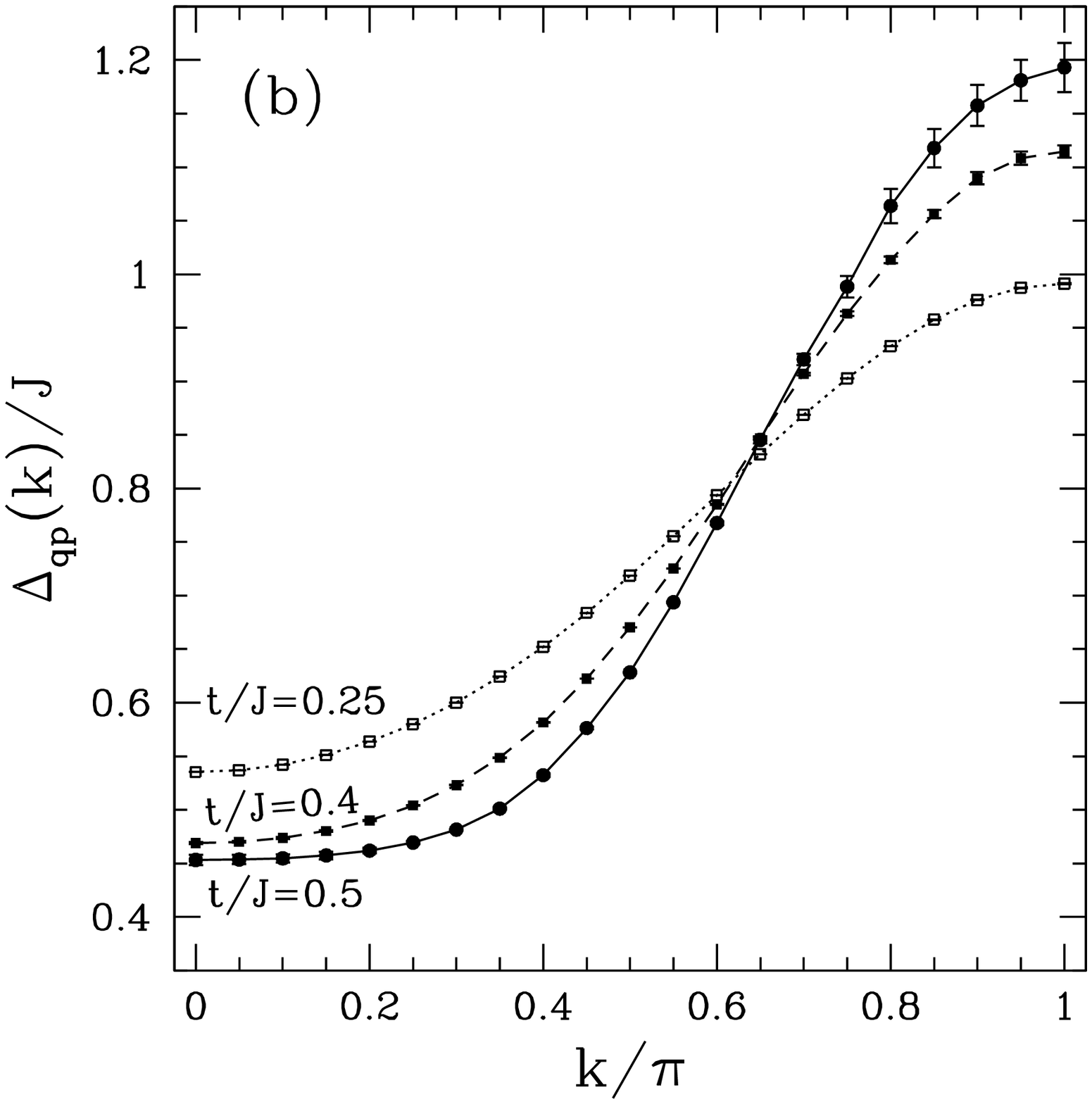}
    \caption{The triplet spin excitation spectrum (a) and 
    one-hole (quasiparticle) excitation spectrum (b)
     for 1D case.
    \label{fig_1d_mk}} 
  }
\end{figure*} 

\begin{figure}[b]
  { \centering
    \includegraphics[width=\figwidth]{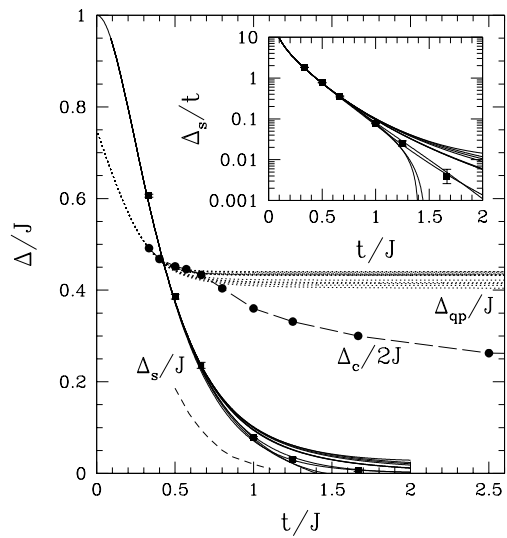}
    \caption{The triplet spin gap $\Delta_s/J$ and quasiparticle gap
    $\Delta_{qp}/J$ versus $t/J$ obtained from
    different  integrated differential approximants.
    The points are the results of DMRG\cite{shi96} for $\Delta_s/J$ (points with
    errorbars) and $\Delta_c/2J$ (solid circles). The
    short dashed line is the result of a mean-field approach\cite{wan93}.
    The inset gives a
    logarithmic plot for $\Delta_s/t$.
    \label{fig_1d_gap}} 
  }
\end{figure} 

Using the dimer expansion approach we have computed series for
the ground state energy in the form
\be
E_0/NJ = \sum_{s=0}^{\infty} e_s (t/J)^s
\ee
and the coefficients, to order 20, are given in Appendix A.
Our coefficients agree exactly with Ref. \onlinecite{shi95},
and add three new terms (odd coefficients vanish for this series).
Integrated differential approximants\cite{gut} are used to evaluated the series
for particular $t/J$, and the resulting energy is shown in Figure
\ref{fig_1d_e0}. The different approximants agree well up to $t/J\simeq 1.2$,
but then splay outwards. We also show, for comparison, the energy obtained
from an early DMRG calculation\cite{yu93}.

Next we turn to the spin excitations. In the strong-coupling limit
a spin excitation corresponds to a spin-triplet at one site, which
is able to propagate coherently via the conduction electron hoping term.
The dispersion relation can be expressed in the form
\be
\Delta_s (k) = \sum_n t_n (\lambda) \cos n k \label{eq4}
\ee
where the quantities $t_n (\lambda)$ are expressed as power series
in $\lambda = t/J$. For the 1D case we have computed these up to
$n =9$ (order 18 in $\lambda$), and for the interested reader we provide 
this data in Appendix A.
Figure \ref{fig_1d_mk}(a) shows the triplet spin-excitation energy vs $k$, for 
value of $t/J=0.25, 0.4, 0.5$. For $t/J=0$ the excitation will, of course,
have energy $J$ and will be dispersionless. Increasing the hopping amplitude
gives increasing bandwidth, with the energy at $k=0$ raised
slightly and a minimum at $k=\pi$. We are unaware of any
previous reported calculations of this dispersion relation,
apart from the second order result given in Ref. \onlinecite{tsu97}.

From eq. \ref{eq4} at $k=\pi$ we obtain a series in $\lambda $ for
the spin gap, which is again evaluated using 
integrated differential approximants.
Results are shown in Figure \ref{fig_1d_gap}. 
The series is well converged up to $t/J\simeq 1.1$.
For comparison we show spin-gaps
calculated by DMRG\cite{shi96} and a mean-field approach\cite{wan93}.
Agreement with DMRG is excellent over the range shown, while the mean-field
method appear to seriously under estimate the size of the gap.
All of the results, including ours, are consistent with a spin-gap which decreases
rapidly but does not vanish until $J=0$.

\begin{figure*}[b]
  { \centering
    \includegraphics[width=\figwidths]{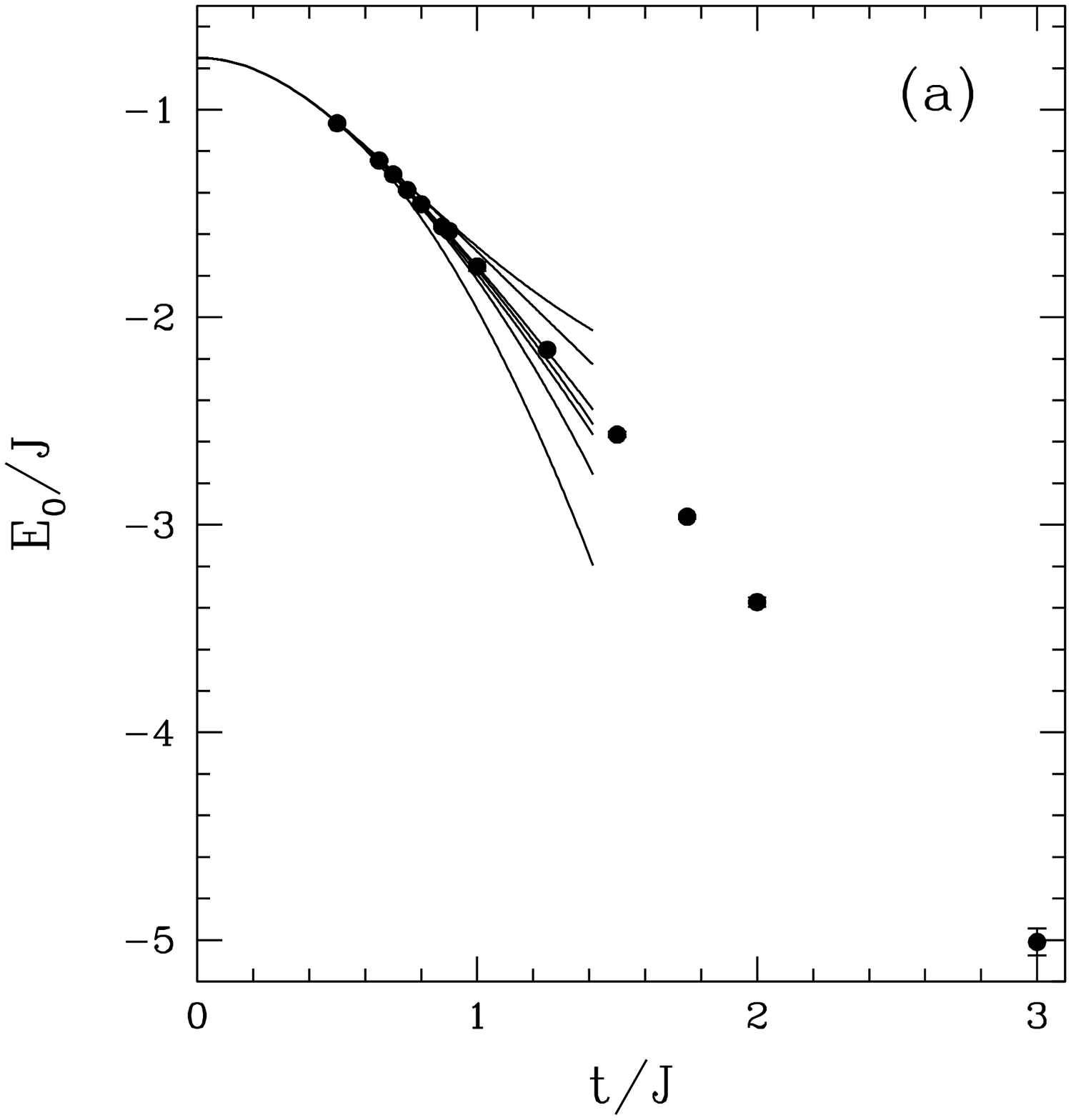}\hspace{1cm}
    \includegraphics[width=\figwidths]{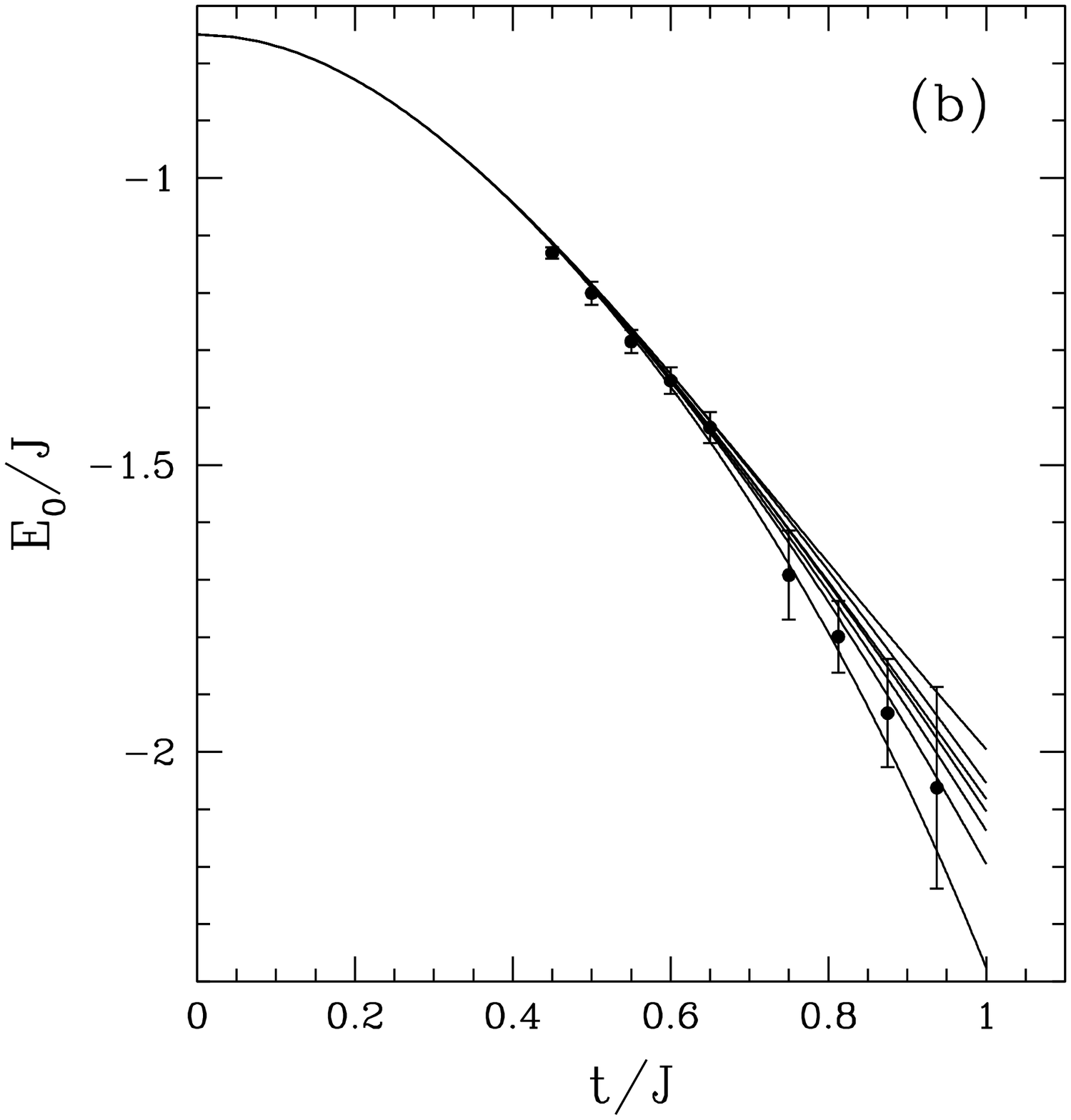} }
    \caption{The ground state energy for the square lattice (a) and
    for simple cubic lattice (b). The solid lines are
     different orders of integrated differential approximants to
      the dimer expansion series, while
    the points with errorbars are the results from Ising expansions.
    \label{fig_2d_e0}} 
\end{figure*} 

Next we consider the so-called ``quasiparticle" excitation, which we prefer to
call a one-hole excitation. This corresponds to the removal of an electron
from the half-filled band and thus, in the strong-coupling limit, to a single
localized spin on one site with singlets on the others. For $t=0$ the energy
gap is thus $3J/4$.  For the 1D case we have computed this series  up to
order 12 in $\lambda$.
Figure \ref{fig_1d_mk}(b) shows the one-hole excitation spectrum for
values $t/J=0.25$, 0.4 and 0.5. The minimum  occurs at $k=0$ and
the bandwidth seems roughly proportional to $t/J$. We are, again, not aware
of any previous calculations of this dispersion curve. The series at $k=0$
allow us to compute the one-hole gap, and our results are plotted 
in Figure \ref{fig_1d_gap}.
A notable feature is that $\Delta_{qp}$ becomes approximately constant
for $t/J> 0.5$. 
We know of no previous calculations for the one-hole gap apart, again, from
the second order strong-coupling result in Ref. \onlinecite{tsu97}.
There is yet another gap, the ``charge gap", which corresponds to an
excitation in which the system remains half-filled, but with a
doubly occupied site and an empty site. We are not able to
compute this via series, at this stage. However, in the strong
coupling limit $\Delta_c = 2 \Delta_{qp}$ (this is valid\cite{tsu97} to 
at least second-order in $t/J$). The charge
gap has been computed by DMRG\cite{shi96}, and we show in Fig. \ref{fig_1d_gap},
the result for $\Delta_c/2$. Evidently for larger hopping parameter
$\Delta_c/\Delta_{qp} < 2$.

\begin{figure*}
  { \centering
    \includegraphics[width=\figwidths]{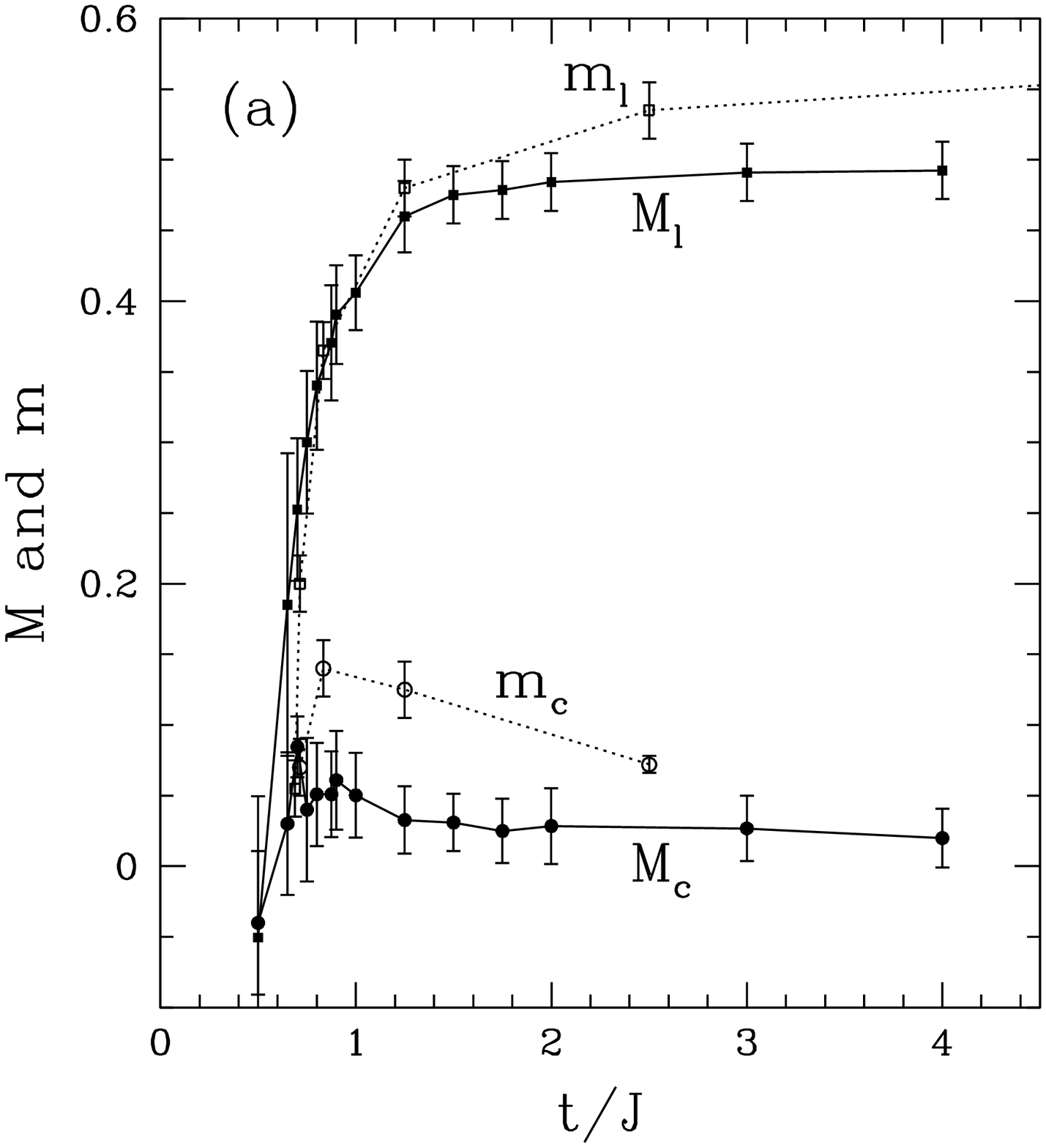}\hspace{1cm}
    \includegraphics[width=\figwidths]{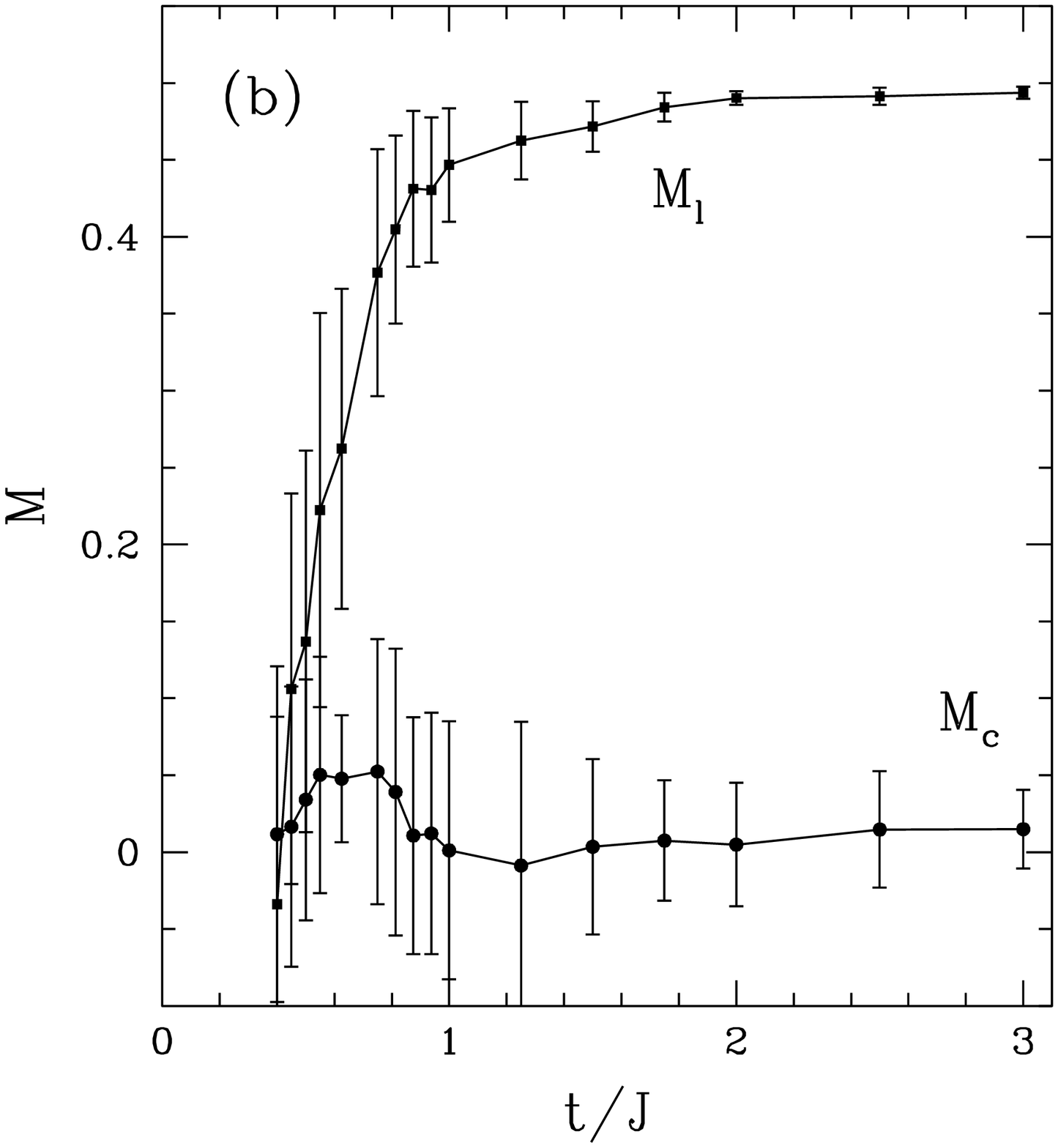} }
    \caption{The staggered magnetizations for both localized
spins ($M_l$) and conduction electrons ($M_c$) for the square lattice (a) and
for simple cubic lattice (b). Also
shown for square lattice are the staggered moments $m$ obtained from 
Monte Carlo calculations\cite{ass99}.
    \label{fig_M_sq}} 
\end{figure*} 

\begin{figure*}
  { \centering
    \includegraphics[width=\figwidths]{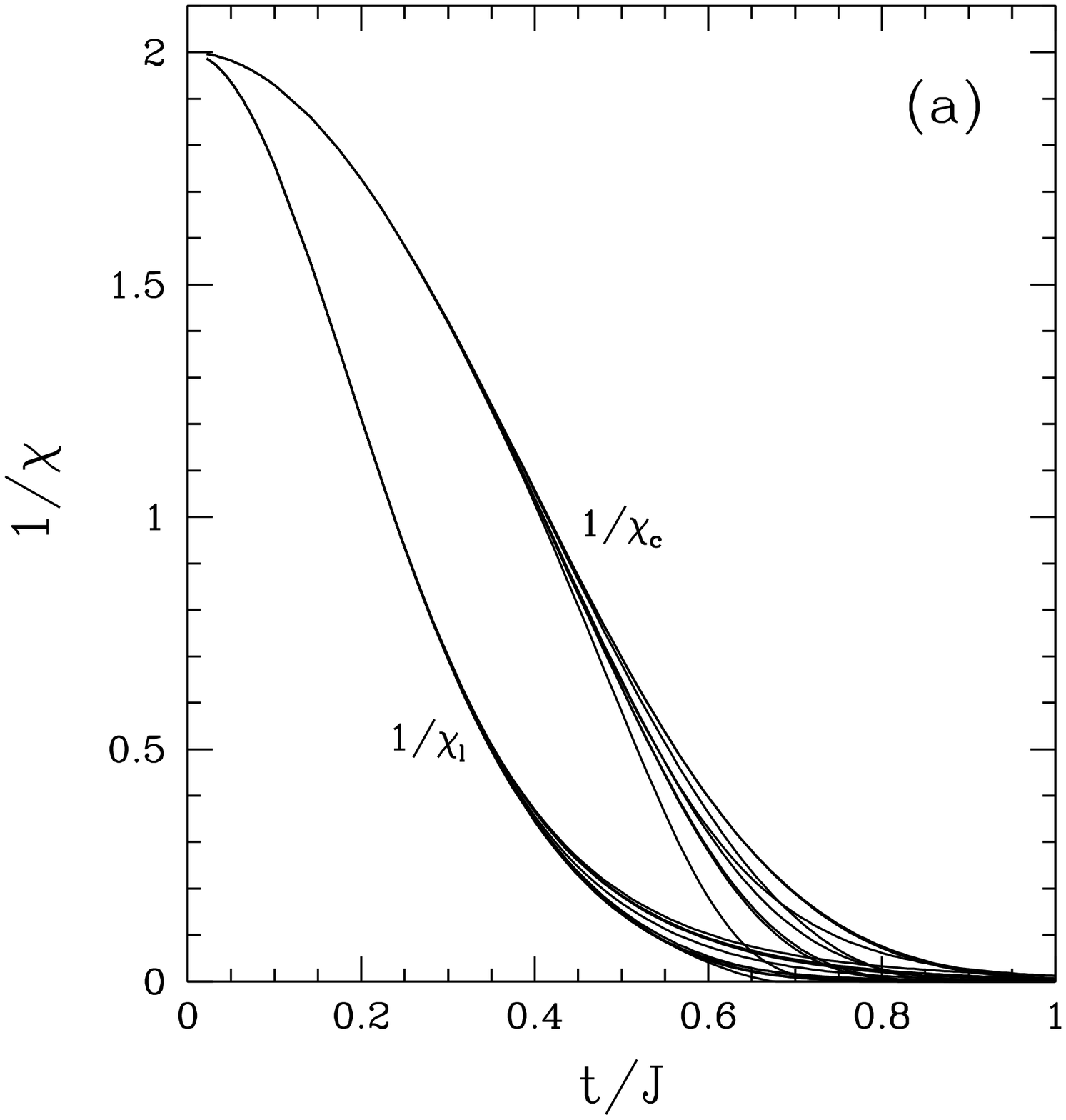}\hspace{1cm}
    \includegraphics[width=\figwidths]{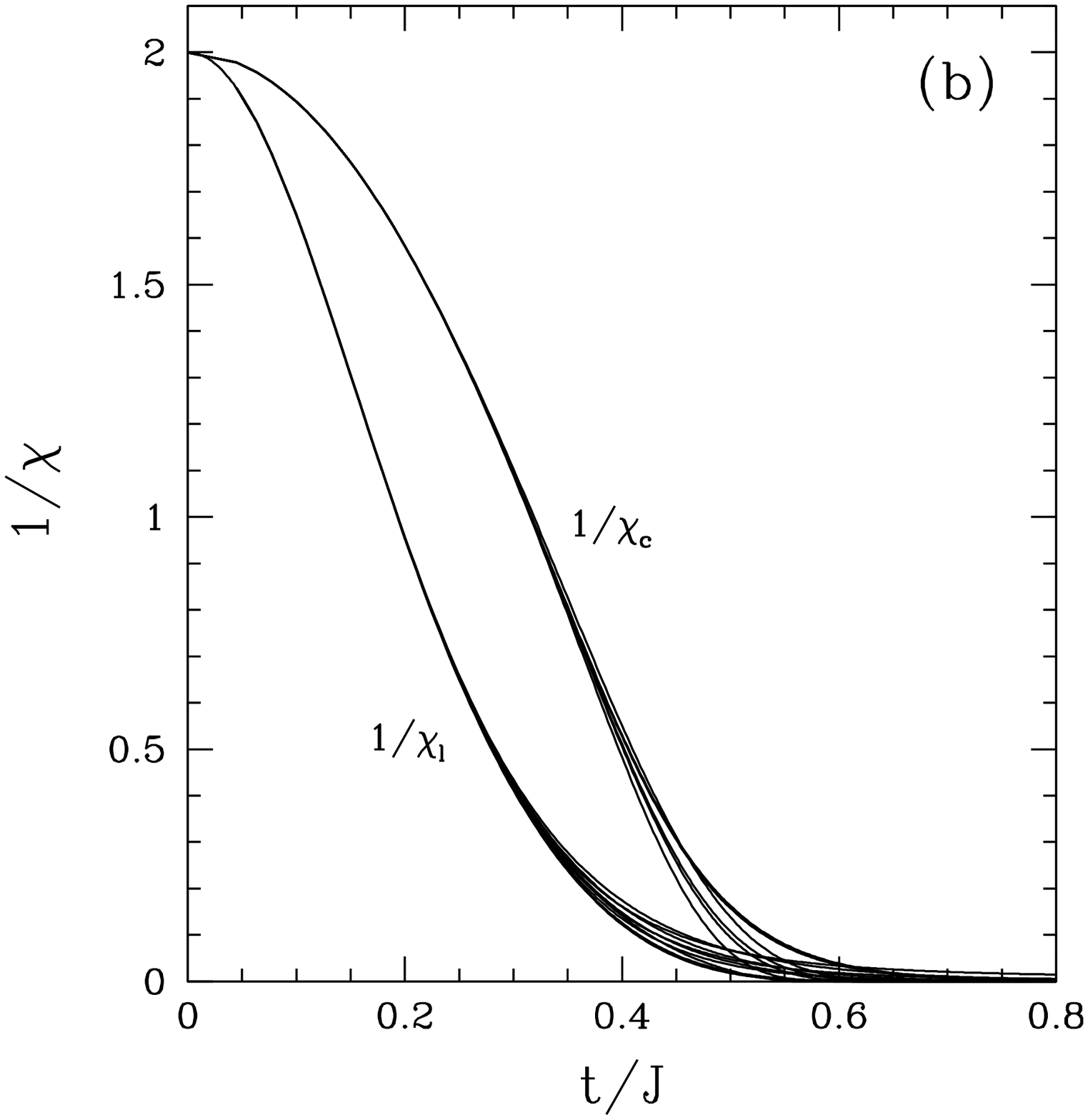}
    }
    \caption{The inverse  antiferromagnetic spin susceptibilities
for both local ($\chi_l$) and intinerant spins ($\chi_c$) for the square lattice (a)
and for simple cubic lattice (b).
    \label{fig_chi_sq}} 
\end{figure*} 

\begin{figure*}
  { \centering
    \includegraphics[width=\figwidths]{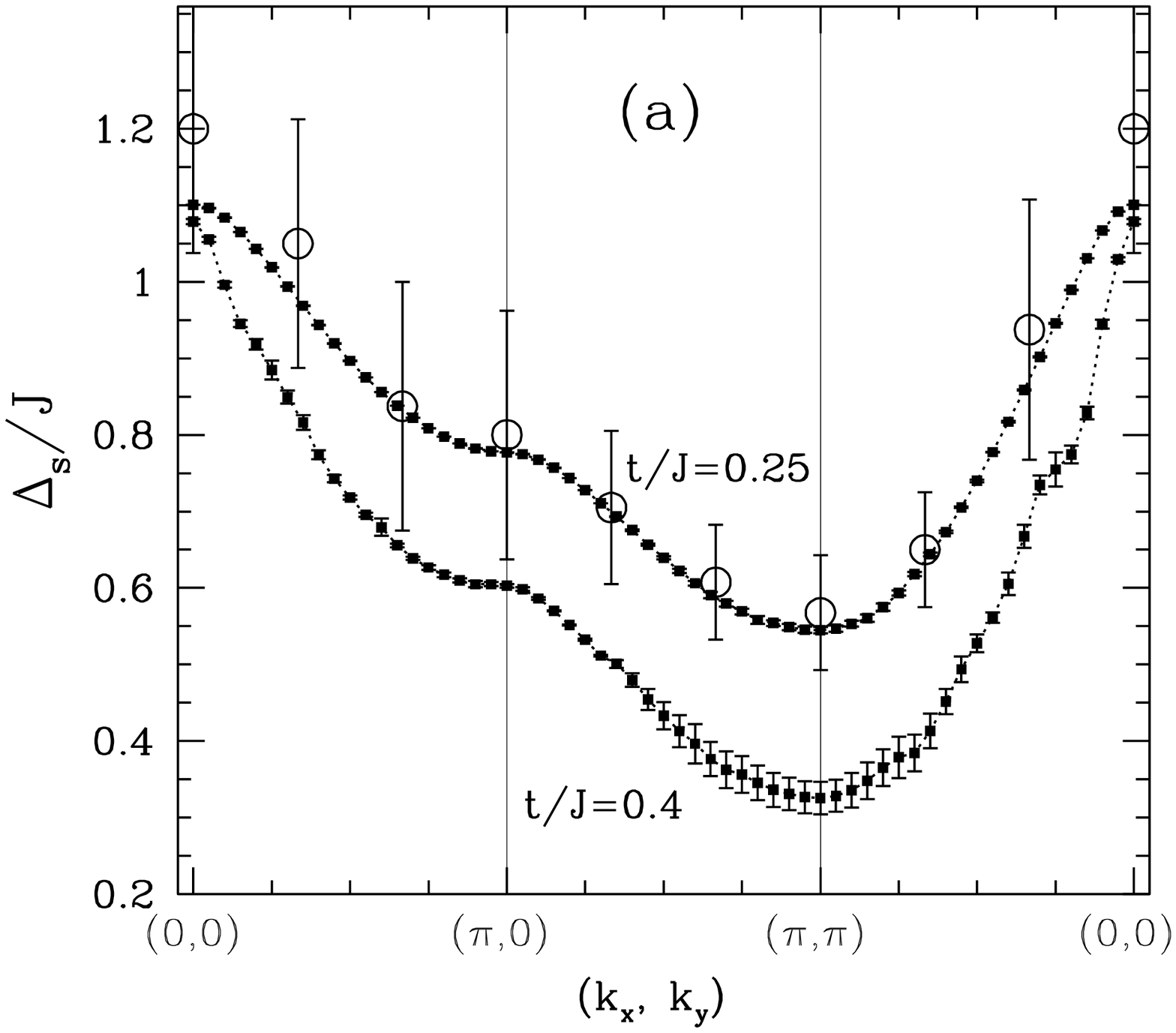}\hspace{1cm}
    \includegraphics[width=\figwidths]{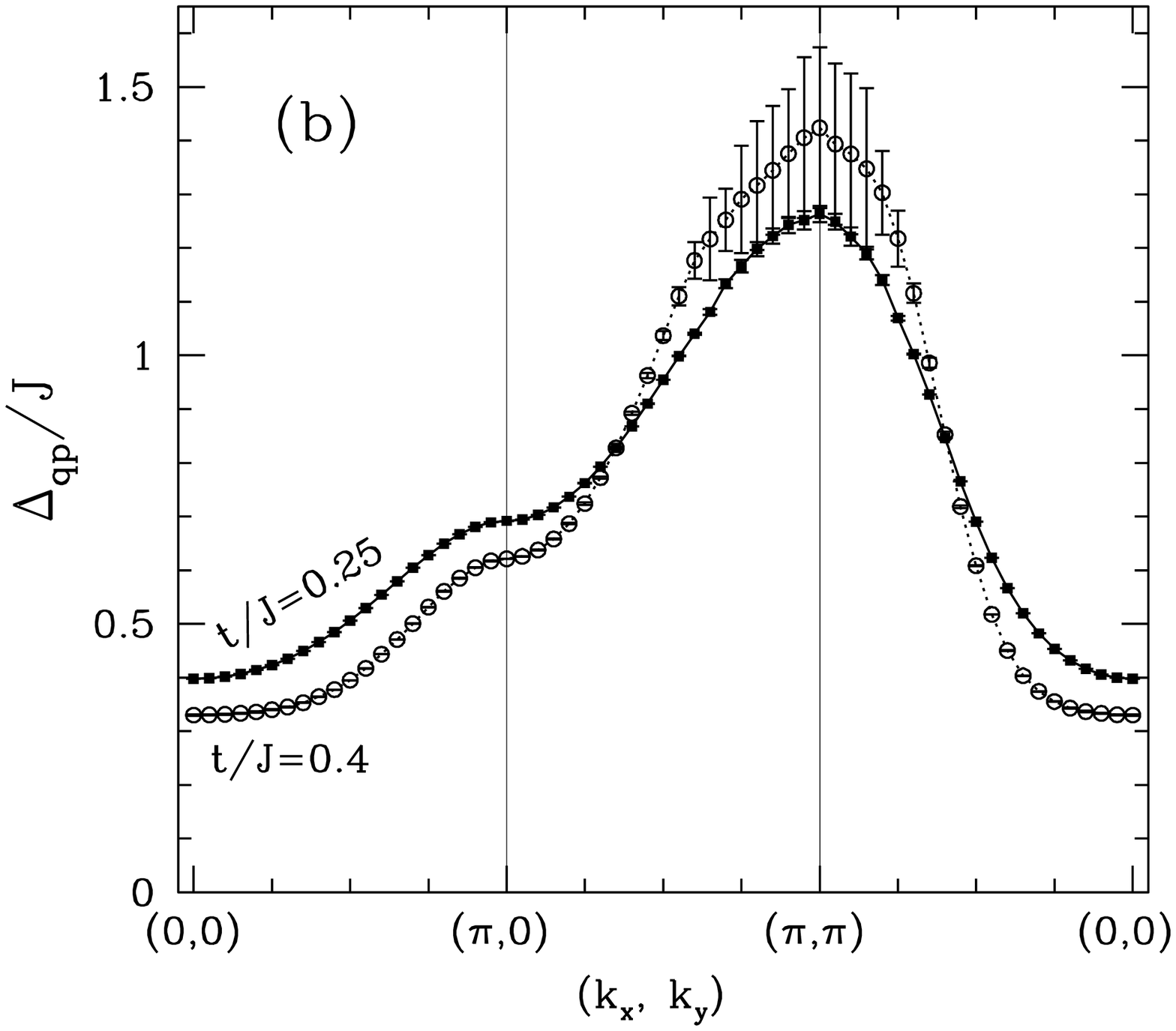}
    }
    \caption{The triplet spin excitation spectrum (a) and quasiparticle excitation spectrum (b)
    for the square lattice. The open points with
    errorbars are variational Monte Carlo results for a $6\times 6$ lattice\cite{wan94}.
    \label{fig_spin_mk_sq}} 
\end{figure*} 

\begin{figure*}
  { \centering
    \includegraphics[width=\figwidths]{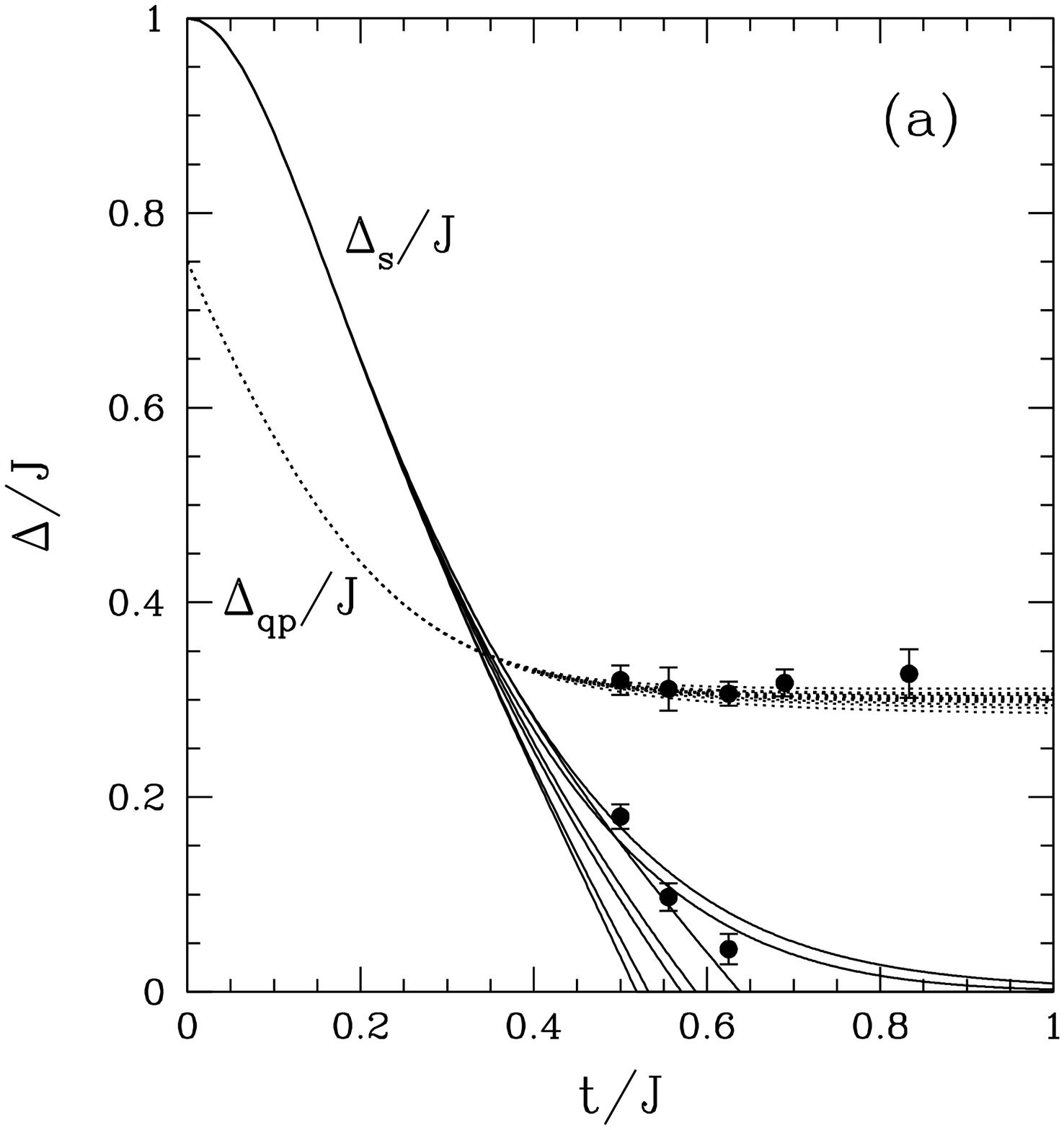}\hspace{1cm}
    \includegraphics[width=\figwidths]{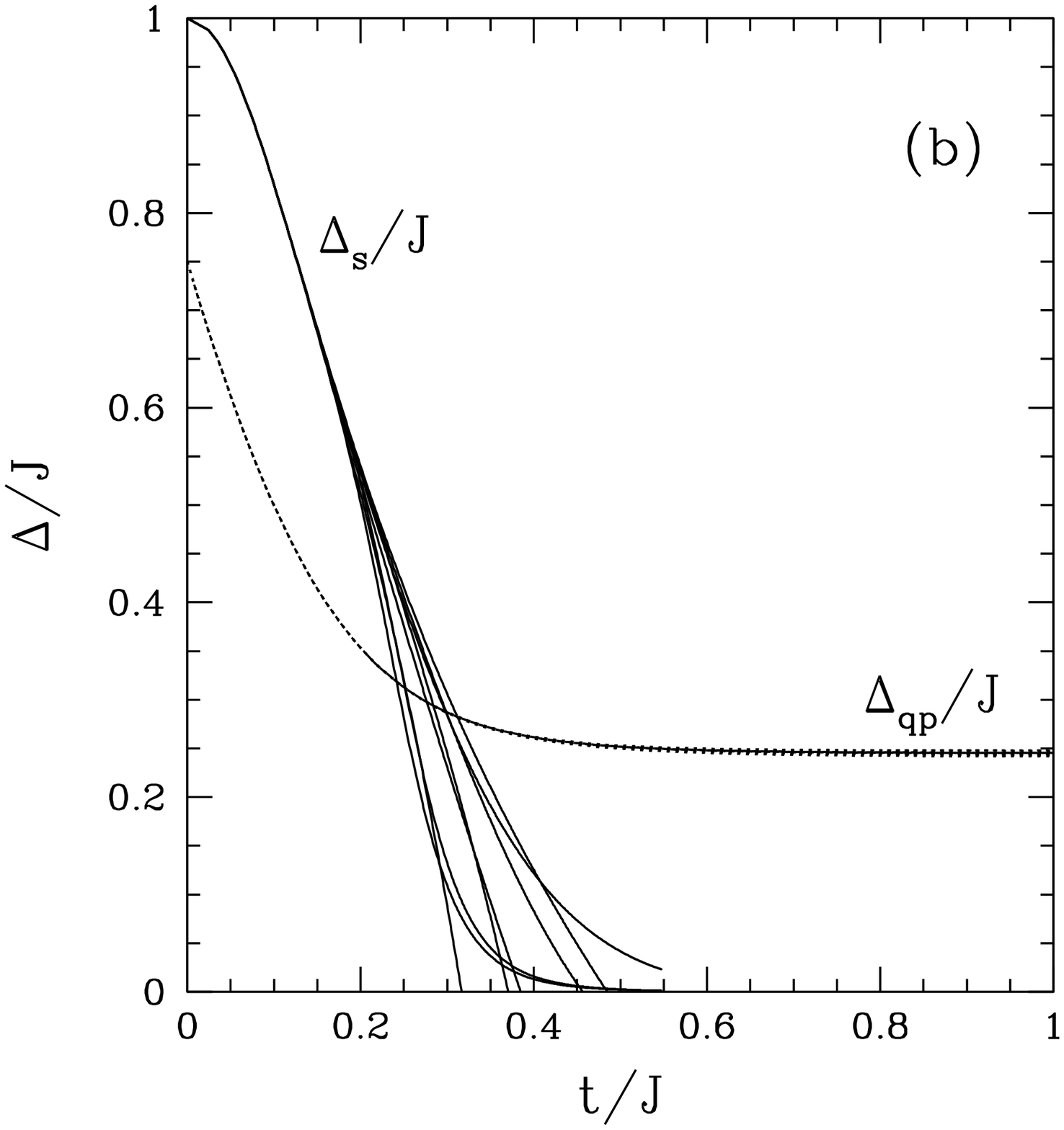}
    }
    \caption{The triplet spin gap $\Delta_s/J$ and quasiparticle gap
    $\Delta_{qp}/J$ for the square lattice (a) and for simple cubic lattice (b)
    obtained from
    different orders of integrated differential approximants.
    The points are the results of a Quantum Monte Carlo study\cite{ass99}.
    \label{fig_2d_gap}} 
\end{figure*} 

\section{\label{sec:2d}The Square and Simple Cubic Lattices}
In two or more spatial dimensions it is believed that the Kondo lattice model
has a true quantum phase transition at some $(J/t)_c$,
between  a gapped spin-liquid
phase and a small $J$ magnetically ordered gapless phase.
For the square lattice quantum Monte Carlo simulations\cite{ass99}
provide strong indications of a transition at $(t/J)_c \simeq 0.69$,
while a bond-operator mean-field theory\cite{jur01} gives
$(t/J)_c = 0.664, 0.546$ for the square and simple cubic lattices,
respectively. A previous series study\cite{shi95} has
given $(t/J)_c \simeq 0.7, 0.5$ respectively. These latter estimates are
relatively imprecise, and it seemed 
worthwhile to investigate this using longer series.

We have derived dimer series for the square lattice for the following 
quantities: ground state energy $E_0$, antiferromagnetic spin susceptibilities
for both local and conduction 
spins ($\chi_l$ and $\chi_c$),  and
the triplet spin excitation spectrum (all to order 12), and the one-hole
(``quasiparticle") excitation spectrum (to order 11).
This adds two non-zero terms to the results of Ref. \onlinecite{shi95}.
The excitation series are new.
In addition we have computed, for the first time, Ising expansions for
the ground state energy and for the staggered  magnetizations (for both local
and itinerant spins) to order 13.
Series, to the same order, have been derived for the simple cubic
lattice for all of the same quantities,
except for the excitations, where we have only computed the minimum gap rather than
the full spectrum.
The dimer series for ground state energy $E_0$, antiferromagnetic 
spin susceptibilities
for both local and conduction  
spins ($\chi_l$ and $\chi_c$),  and
the minimum triplet spin gap are given in Table I. 
Our results agree completely with those of Ref. \onlinecite{shi95}
for the square lattice but disagree for the simple cubic lattice 
susceptibilities beyond the 4th term.  We are unable to resolve this,
but we confident that our results are correct.

We first show, in Figure \ref{fig_2d_e0}, our estimate of the ground state
energy, as a function of $t$, obtained from both dimer
and Ising expansions. Both series converge well for small $t$, but the
Ising expansion has better convergence for larger $t$. There are no
previously reported data for $E_0$. There is no evidence of any
anomaly in $E_0$ at the supposed phase transition point, nor
would we expect this.

The most direct way of identifying any critical point $(t/J)_c$
is from poles of Dlog Pad\'e approximants\cite{gut}. However the
series are irregular (Table I) and, perhaps not surprisingly,
this yields no consistent results. However if we know, or
assume, the value of the critical exponent $\gamma$ then biased
estimates of the critical point can be obtained from direct 
Pad\'e approximants to the series for $\chi^{1/\gamma}$,
which should have a simple pole.
Here we expect the transition for the square lattice to be of the same
universality class as the $d=3$ classical Heisenberg model with
$\gamma\simeq 1.4$, and for the simple cubic lattice to be in
the $d=4$ universality class, with $\gamma=1.0$. In Table II
we show estimates of $x_c^2$ $(x=t/J)$ obtained in this way. As
can be seen, a number of consistent estimates of the pole
are obtained, particularly for the series for the conduction
electron susceptibility. We might reasonably estimate
for the square lattice $(t/J)_c = 0.68\pm 0.02$,
and for the simple cubic lattice $(t/J)_c = 0.46\pm 0.01$,
where the error are subjective confidence limits.
These correspond to $(J/t)_c = 1.48$, 2.15 for the 2D and
3D case, values which are consistent with previous
estimates.


An alternative approach is to evaluate the staggered magnetization and 
susceptibility directly  via integrated differential approximants and
to look at the behaviour as a function of $t/J$. 
The staggered magnetizations shown
in Figure \ref{fig_M_sq} are obtained from an Ising expansions,
starting from an antiferromagnetically ordered state. These are
relatively  constant for $t/J>1$, but drop sharply to zero around
$t/J \sim 0.6-0.7$ in 2D and $t/J \sim 0.5$ in 3D, confirming the existence of a transition to a
magnetically disordered phase. The error limits are rather large, 
and it is not possible to determine the transition point with high
precision in this way. Our  magnetization curves in 2D are very similar to 
the 
Quantum Monte Carlo results\cite{ass99},  although our
conduction election magnetization is much smaller. We show the
QMC results for comparison.
In Figure \ref{fig_chi_sq} we show curves of the inverse susceptibilities,
in the Kondo phase, obtained from dimer expansions. Again there is
clear evidence for a transition, but it is difficult to locate
precisely. Our best estimates from these figures would be $t/J \simeq 0.75\pm 0.10$ in
2D and $t/J\simeq 0.50\pm 0.05 $ in 3D, less
precise but quite consistent with the direct Pad\'e approximant estimates.

Next we consider the spin triplet and 1-hole (quasiparticle) excitations.
Figure \ref{fig_spin_mk_sq}(a) shows the full spin triplet dispersion curve
for $t/J = 0.25, 0.4$, both in the Kondo phase, for 2D. For comparison we also
show variational Monte Carlo results of Wang {\it et al.}\cite{wan94}.
The lowest spin excitation is at $(\pi,\pi)$, and the spin gap clearly
decreases as $t$ increases.

In Figure \ref{fig_spin_mk_sq}(b) we show the dispersion relation for
the 1-hole excitation, again for $t/J =0.25, 0.4$, for the 2D case. The minimum occurs
at ${\bf k}= (0,0)$, with a maximum at $(\pi,\pi)$. The overall shape
is qualitatively similar to the mean-field results of Ref. \onlinecite{jur01}.
We note the large error bars near $(\pi,\pi)$ for
$t/J=0.4$. This may be a signature of degeneracy with a 2-particle continuum,
as was observed in the $J_1-J_2$ model\cite{j1j2}.

Series have also been computed directly for the spin gap and quasiparticle gap, for
both 2D and 3D cases. Analysis is shown in Figure 8. The data clearly
show the spin gap decreasing to zero at a critical point
$(t/J)_c$, the position being consistent with our
estimates above. For the square lattice our results are in excellent agreement
with the QMC results. For the 3D lattice we are unaware of any previous results
for either energy gap.

\section{\label{sec:con}Summary and Discussion}
We have used linked cluster series methods in a comprehensive study
of zero-temperature properties of the Kondo lattice model at half-filling,
for the linear chain, square lattice, and simple cubic lattice. Our work
significantly extends a previous series study\cite{shi95} for the
ground state energy and susceptibility, and presents new series results for
the magnetization, spin and quasiparticle dispersion relations, and energy gaps.
Wherever possible we have compared our results with calculations by other methods and,
in general, find excellent agreement.

Our analysis supports the existence of a quantum critical point in the 2D and 3D cases,
separating a Kondo spin-liquid phase (large $J$) from an ordered phase.
Our estimates for the critical point are
$(t/J)=0.68\pm 0.02$, $0.46\pm 0.01$ in 2D and 3D respectively. 
These results are from direct Pad\'e approximants to the series for
$\chi^{1/\gamma}$, and hence are biassed by, but not particularly sensitive to,
the choice of $\gamma$. The critical point estimate in 2D agrees very well
with previous estimates, while in 3D we find a slightly higher value of
$(J/t)_c$, 2.17, compared with 1.833 from mean-field.

We have not yet analysed the data for the frustrated close-packed triangular
and face-centered cubic lattices, or for the body-centered cubic lattice, which we also
have computed. Nor have we explored the possible existence of bound states, or the region away
from half-filling. Recent developments in series methods\cite{zwh1,zwh2} make this possible, 
and we intend to pursue these directions, as well as others, in future work.

\appendix
\subsection*{Appendix}

The dimer series of ground state energy for 1D KLM are:
\begin{widetext}
\bea
E_0/NJ &=& -3/4 -2/3 \lambda^ 2 -14/45 \lambda^4 
 +0.71146854791 \lambda^6   
 +0.19256921565 \lambda^8   \nonumber \\
&& -2.8528410569 \lambda^{10}   
 +2.2809484235 \lambda^{12}   
 +12.882850521 \lambda^{14}   
 -30.446998097 \lambda^{16}  \nonumber \\
 && -43.303667770 \lambda^{18}   
 +270.26259775 \lambda^{20}   + O(\lambda^{22})
\eea

The triplet spin excitation spectrum for 1D KLM are
\bea
\Delta_s (k)&&/J =
  1 - {\frac{8\,{{\lambda}^2}}{3}} + {\frac{2356\,{{\lambda}^4}}{135}} - 
   181.93078\,{{\lambda}^6} + 2152.5066\,{{\lambda}^8} 
   -27691.038\,{{\lambda}^{10}} + 329428.5445\,{{\lambda}^{12}} \nonumber \\ 
    && - 
   2.5917002\,{{10}^6}\,{{\lambda}^{14}} - 2.75912\,{{10}^7}\,{{\lambda}^{16}} + 
   2.23129\,{{10}^9}\,{{\lambda}^{18}} 
    +   {\Big (} 4\,{{\lambda}^2} - {\frac{272\,{{\lambda}^4}}{9}} + 
      305.675\,{{\lambda}^6} - 4076.18\,{{\lambda}^8} \nonumber \\
   &&   + 
      55158.531287\,{{\lambda}^{10}} - 694363.\,{{\lambda}^{12}} 
      +  6.4442\,{{10}^6}\,{{\lambda}^{14}} + 
      2.20385\,{{10}^7}\,{{\lambda}^{16}} - 
      3.67419\,{{10}^9}\,{{\lambda}^{18}} {\Big )} \,\cos (k) \nonumber \\
     && + 
   {\Big (} {\frac{308\,{{\lambda}^4}}{27}} - 203.122\,{{\lambda}^6} + 
      3199.476135\,{{\lambda}^8} - 50109.7\,{{\lambda}^{10}} + 
      729779.9656\,{{\lambda}^{12}} - 
      8.92778\,{{10}^6}\,{{\lambda}^{14}} \nonumber \\
    && +  5.42326\,{{10}^7}\,{{\lambda}^{16}} + 
      1.72156\,{{10}^9}\,{{\lambda}^{18}} {\Big )} \,\cos (2\,k) 
    +    {\Big (} 71.0803\,{{\lambda}^6} - 1809.21\,{{\lambda}^8} + 
      35912.6434\,{{\lambda}^{10}} \nonumber \\
      && - 640789.\,{{\lambda}^{12}} + 
      1.0026\,{{10}^7}\,{{\lambda}^{14}} - 
      1.22082\,{{10}^8}\,{{\lambda}^{16}} + 4.1027\,{{10}^8}\,{{\lambda}^{18}}
       {\Big )} \,\cos (3\,k) 
      + 
   {\Big (} 521.4000489\,{{\lambda}^8} \nonumber \\
   && - 17575.\,{{\lambda}^{10}} + 
      416938.0779\,{{\lambda}^{12}} - 
      8.34562\,{{10}^6}\,{{\lambda}^{14}} + 
      1.40949\,{{10}^8}\,{{\lambda}^{16}} - 
      1.77087\,{{10}^9}\,{{\lambda}^{18}} {\Big )} \,\cos (4\,k) \nonumber \\
    &&  +
    \left( 4305.106\,{{\lambda}^{10}} - 178962.\,{{\lambda}^{12}} + 
      4.91952\,{{10}^6}\,{{\lambda}^{14}} - 
      1.09405\,{{10}^8}\,{{\lambda}^{16}} + 
      2.00128\,{{10}^9}\,{{\lambda}^{18}} \right) \,\cos (5\,k) \nonumber \\
    &&  +
    \left( 38024.843\,{{\lambda}^{12}} - 
      1.87593\,{{10}^6}\,{{\lambda}^{14}} + 
      5.85515\,{{10}^7}\,{{\lambda}^{16}} - 
      1.43507\,{{10}^9}\,{{\lambda}^{18}} \right) \,\cos (6\,k) \nonumber \\
    &&  +
    \left( 351790.5722\,{{\lambda}^{14}} - 
      2.00559\,{{10}^7}\,{{\lambda}^{16}} + 
      7.00211\,{{10}^8}\,{{\lambda}^{18}} \right) \,\cos (7\,k) \nonumber \\
    &&  +
    \left( 3.36497\,{{10}^6}\,{{\lambda}^{16}} - 
      2.17486\,{{10}^8}\,{{\lambda}^{18}} \right) \,\cos (8\,k) +
    3.30089\,{{10}^7}\,{{\lambda}^{18}}\,\cos (9\,k)
\eea
\end{widetext}


\begin{acknowledgments}
This work  forms part of 
a research project supported by a grant
from the Australian Research Council.
The computations were performed on an AlphaServer SC
 computer. We are grateful for the computing resources provided
 by the Australian Partnership for Advanced Computing (APAC)
National Facility.
\end{acknowledgments}


\newpage
\bibliography{basename of .bib file}


\begin{table*}
\squeezetable
\caption{Series coefficients for dimer expansions for the ground energy per site $E_0/JN$, 
the minimum triplet spin gap $\Delta_s/J$, and the antiferromagnetic 
spin susceptibilities
for both local and conduction  
spins ($\chi_l$ and $\chi_c$).
Nonzero coefficients $(t/J)^n$
up to order $n=12$ for square lattice and simple cubic lattice
 are listed.}\label{tab_ser}
\begin{ruledtabular}
\begin{tabular}{rllll} 
\multicolumn{1}{c}{$n$} &\multicolumn{1}{c}{$E_0/JN$} 
&\multicolumn{1}{c}{$\Delta_s/J$}
&\multicolumn{1}{c}{$\chi_c$}  &\multicolumn{1}{c}{$\chi_l$} \\
\hline
\multicolumn{5}{c}{square lattice}\\
 0 & -7.500000000$\times 10^{-1}$ & ~1.000000000                & ~5.000000000$\times 10^{-1}$ & ~5.000000000$\times 10^{-1}$ \\
 2 & -1.333333333                 & -1.333333333$\times 10^{1}$ & ~1.777777778                 & ~6.518518519       \\
 4 & -8.888888889$\times 10^{-2}$ & ~1.611851852$\times 10^{2}$ & ~4.632888889                 & ~3.850916872$\times 10^{1}$ \\
 6 & ~4.231487360                 & -2.640795767$\times 10^{2}$ & ~6.236968731                 & ~4.067531551$\times 10^{1}$ \\
 8 & -1.519899530$\times 10^{1}$  & -1.072124632$\times 10^{5}$ & ~4.340134129$\times 10^{1}$  & ~6.687413797$\times 10^{1}$ \\
10 & ~8.900723089$\times 10^{-1}$ & ~5.495699137$\times 10^{6}$ & ~1.004232541$\times 10^{2}$  & ~1.823289007$\times 10^{3}$ \\
12 & ~3.109971534$\times 10^{2}$  & -1.561259011$\times 10^{8}$ & -6.384897515$\times 10^{2}$  & -4.654864554$\times 10^{3}$ \\
\hline
\multicolumn{5}{c}{simple cubic lattice}\\
  0 & -7.500000000$\times 10^{-1}$ & ~1.000000000                & ~5.000000000$\times 10^{-1}$ & ~5.000000000$\times 10^{-1}$ \\
  2 & -2.000000000                 & -2.000000000$\times 10^{1}$ & ~2.666666667                 & ~9.777777778       \\
  4 & ~6.666666667$\times 10^{-1}$ & ~3.063111111$\times 10^{2}$ & ~1.291555556$\times 10^{1}$  & ~8.443753086$\times 10^{1}$ \\
  6 & ~7.834807760                 & ~1.060237771$\times 10^{3}$ & ~5.132587514$\times 10^{1}$  & ~2.399390035$\times 10^{2}$ \\
  8 & -7.114791245$\times 10^{1}$  & -2.760658123$\times 10^{5}$ & ~3.510946982$\times 10^{2}$  & ~1.684588635$\times 10^{3}$ \\
 10 & ~3.498368486$\times 10^{2}$  & ~3.860398011$\times 10^{6}$ & ~1.354232378$\times 10^{3}$  & ~1.217247537$\times 10^{4}$ \\
 12 & -1.603723348$\times 10^{2}$  & ~5.557190900$\times 10^{8}$ & ~2.797156237$\times 10^{3}$  & -2.514875289$\times 10^{4}$ \\
\end{tabular}                                                       
\end{ruledtabular}                                                  
\end{table*}

\begin{table*}
\caption{Estimates of $x_c^2 = (t/J)^2_c$ from poles of $[N,D]$ Pad\'e approximants to the series
for $\chi_{c,l}^{1/\gamma}$. The index $c,l$ denotes the series for conduction electron,
localized spin respectively.}\label{tab_pade}
\begin{ruledtabular}
\begin{tabular}{ccccc} 
\multicolumn{1}{c}{\underline{Square Lattice} } 
  & \multicolumn{1}{c}{Approximant} & \multicolumn{1}{c}{$x_c^2$} 
  & \multicolumn{1}{c}{Approximant} & \multicolumn{1}{c}{$x_c^2$}\\
 $\gamma=1.4$ & $[2,3]_c$ & 0.45378    &  $[3,3]_c$ & 0.46298  \\
               & $[3,3]_l$ & 0.46990    &  $[2,4]_c$ & 0.46350  \\
\multicolumn{5}{c}{ Estimate $x_c^2 =0.46\pm 0.02$, $(t/J)_c =0.68\pm 0.02$ } \\
\hline
\multicolumn{1}{c}{\underline{Simple Cubic Lattice} } 
  & \multicolumn{1}{c}{Approximant} & \multicolumn{1}{c}{$x_c^2$} 
  & \multicolumn{1}{c}{Approximant} & \multicolumn{1}{c}{$x_c^2$}\\
 $\gamma=1.0$  & $[1,1]_c$ & 0.20647    &  $[0,2]_c$ & 0.20888  \\
               & $[2,2]_c$ & 0.21660    &  $[2,2]_l$ & 0.21637  \\
               & $[1,3]_c$ & 0.22094    &  $[3,2]_c$ & 0.19300  \\
               & $[2,3]_c$ & 0.20601    &  $[4,2]_c$ & 0.19480  \\
               & $[3,3]_c$ & 0.20798    &  $[3,3]_l$ & 0.20078  \\
               & $[2,4]_c$ & 0.20809    &                       \\
\multicolumn{5}{c}{ Estimate $x_c^2 =0.21\pm 0.01$, $(t/J)_c =0.46\pm 0.01$ } \\
\end{tabular}                                                       
\end{ruledtabular}                                                  
\end{table*}

\end{document}